\documentclass[showpacs,preprintnumbers,amsmath,amssymb,twocolumn]{revtex4}
\usepackage{graphicx}
\usepackage{dcolumn}
\usepackage{bm}
\usepackage{amssymb}
\usepackage{array}
\usepackage{longtable}

\begin{document}

\title{Random walks in modular scale-free networks with multiple traps}

\author{Zhongzhi Zhang}
\email{zhangzz@fudan.edu.cn}

\author{Yihang Yang}

\author{Yuan Lin}

\affiliation {School of Computer Science, Fudan University,
Shanghai 200433, China}

\affiliation {Shanghai Key Lab of Intelligent Information
Processing, Fudan University, Shanghai 200433, China}

\begin{abstract}
Extensive empirical investigation has shown that a plethora of real
networks synchronously exhibit scale-free and modular structure, and
it is thus of great importance to uncover the effects of these two
striking properties on various dynamical processes occurring on such
networks. In this paper, we examine two cases of random walks performed on a
class of modular scale-free networks with multiple traps located at
several given nodes. 
We first derive a formula of the mean first-passage time (MFPT) for a general network, which is the mean
of the expected time to absorption originating from a specific node,
averaged over all non-trap starting nodes. Although the computation is complex, the expression of the formula is exact; moreover, the computational approach and procedure are independent of the number and position of the traps. We then determine analytically the MFPT for the two random walks being considered. The obtained analytical results are in complete agreement with the numerical ones. Our results show
that the number and location of traps play an important role in the behavior of the MFPT, since for both cases the MFPT grows as a power-law function of the number of nodes,
but their exponents are quite different. We demonstrate that the root of the
difference in the behavior of MFPT is attributed to the modular and scale-free
topologies of the networks. This work can deepen the understanding of diffusion on networks with
modular and scale-free architecture and motivate relevant studies
for random walks running on complex random networks with multiple
traps.
\end{abstract}



\date{\today}
\maketitle

\section{Introduction}

In the past decade, with a huge amount of data and computational
resources available, scientists have processed and analyzed data of
a wide variety of real systems in different areas, leading to
important advances in the understanding of complex
systems~\cite{AlBa02,DoMe02,Ne03,BoLaMoChHw06}. A large volume of
empirical studies showed that scale-free feature~\cite{BaAl99} and
modular structure~\cite{GiNe02,RaSoMoOlBa02,RaBa03} are two
prominent properties that seem to be common to real networks,
especially biological and social networks. The former implies that
the networks obey a power-law degree distribution as $P(k) \sim
k^{-\gamma}$ with $2< \gamma \leq 3$, while the latter means that
the networks can be divided into groups (modules), within which
nodes are more tightly connected with each other than with nodes
outside. These two remarkable natures constitute our fundamental
understanding of the structure of complex networks, which are
relevant to other topological features (i.e., average
distance~\cite{ChLu02,CoHa03} and clustering
coefficient~\cite{RaBa03}), and have led to many popular topics of
research in network science, including explaining the origin of
the scale-free phenomenon~\cite{AlBa02,DoMe02}, identifying the
modules~\cite{DaDuDiAr05,PaDeFaVi05,Ne06,Fo10,ZhZhZhGu11} and finding their
concrete applications~\cite{Pe01,WuGaHa04}.

It is well known that one of the ultimate goals for research on
complex networks is to make clear how the underlying structural
characteristics affect the dynamical processes defined on
networks~\cite{Ne03,DoGoMe08}. Among various dynamical processes,
random walks have held continual interest within the scientific
community~\cite{HaBe87,MeKl00,MeKl04,BuCa05,SoMaBl97,PaAm04,NoRi04,BeCoMoSuVo05,SoRebe05,Bobe05,GaSoHaMa07,BaBeWi08,BaCaPa08,KiCaHaAr08,ZhZhZhYiGu09,HaRo09,ZhAlHoZhCh11}
because of their relevance to a wide range of different applications
to many fields~\cite{We94,Hu95}. In particular, as an integral
subject of random walks, the trapping problem is closely related to
numerous aspects in a great many
disciplines~\cite{JaBl01,Sh05,WhGo99,FoPiReSa07}. Over the past
decades, scholars in a large interdisciplinary community have made a
huge effort to address the trapping problem in diverse networks,
including regular lattices~\cite{Mo69}, regular
fractals~\cite{KaBa02PRE,KaBa02IJBC,
ZhWuZhZhGuWa10,Ag08,ZhLiZhWuGu09}, small-world
networks~\cite{CaAb08}, and scale-free
networks~\cite{ZhQiZhXiGu09,ZhZhXiChLiGu09,ZhGuXiQiZh09,AgBu09,TeBeVo09,ZhXiZhGaGu09,ZhXiZhLiGu09},
among other graphs~\cite{CoBeTeVoKl07,BeMeTeVo08,CoTeVoBeKl08}.

Thus far, most previous works on random walks in complex networks have
focused on the case with a single trap fixed at a given location,
while work on the case with multiple traps is much less common.
In particular, research on the multiple-trap problem in complex
networks with modular organization and scale-free structure is still
lacking, despite the multiple-trap issue having obvious
applications to various aspects~\cite{YuAc03} (description of
particle-cluster aggregation~\cite{WiSa82,WiSa83}, for instance) and
being relevant in diffusion-limited reactions in chemical
field~\cite{Ri85} and modular and scale-free topologies having
vital influence on dynamical processes taking place on
networks~\cite{Hi07,Ma07,Di08,ZhLiGoZhGuLi09,BuSp09}. ¡¡

In this paper, we study the classic random-walk problem for a
category of modular scale-free networks~\cite{RaSoMoOlBa02,RaBa03}
with several given nodes being occupied by immobile traps, which absorb
all particles visiting it. The basic quantity we are interested in
is the mean first-passage time (MFPT)~\cite{Re01} characterizing the
trapping process, which is defined as the average of expected time
for a particle starting off from a particular node until first
visiting one of the traps, averaged over all nontrap source nodes.
The networks we study are of a deterministic family, which has
proven to be an important tool in the field of complex networks and
has recently attracted much
interest~\cite{DoGoMe02,JuKiKa02,CoOzPe00,ZhRoGo06,HiBe06,RoHaAv07,ZhZhChGu07,ZhZhFaGuZh07,BoGoGu08,CoMi09,MoRoDa09,ZhGuDiChZh09,MiCoChZh10}.

We first study the MFPT in a generic network and reduce the
problem of computing the MFPT to finding the sum of elements of a
matrix associated with the trapping issue. Then, based on the deterministic recursive construction of the modular scale-free networks being considered, we investigate analytically the key of quantity MFPT for two cases of particular arrangements of traps. In the first case, traps are placed on peripheral nodes; in the other case, traps are fixed on those nodes farthest from the hub. For both cases, we derive exactly the dominant scalings for the MFPT and show that they produce a power-law function of the network size with their exponents smaller than 1 but being different, which is confirmed by the numerical results obtained via inverting related matrices. The obtained results indicate that both the number and the location of traps have a significant impact on the behavior of the trapping. We demonstrate that the high
efficiency of both trapping processes and the distinction between the behavior of the MFPT for the two random walks are rested with the
scale-free property and the modular structure of the networks under consideration.

\section{Modular scale-free networks}

\begin{figure}
\begin{center}
\includegraphics[width=0.8\linewidth,trim=0 0 0 -3]{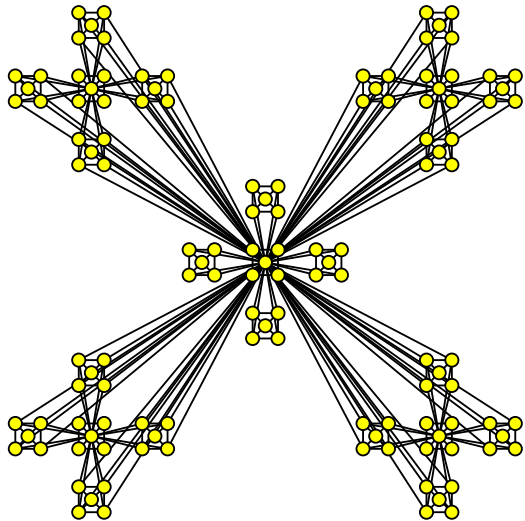}
\end{center}
\caption[kurzform]{(Color online) Sketch of a network $M_3$ for the
limiting case of $m=5$. Note that the diagonal nodes are also linked to each other; the edges are not visible.} \label{network}
\end{figure}

We first introduce the model for the modular scale-free networks,
which are built in an iterative way~\cite{RaSoMoOlBa02,RaBa03}. Let
$M_{g}$ stand for the network model after $g$ ($g\geq 1$) iterations
(i.e., number of generations). Initially ($g=1$), the model is
composed of $m$ ($m \geq 3$) nodes linked by $m(m-1)/2$ edges
forming a complete graph, among which a node (e.g., the central node
in Fig.~\ref{network}) is called hub (or root) node, and the other
$m-1$ nodes are named peripheral nodes. At the second generation
($g=2$), $m-1$ replicas of $M_{1}$ are created with the $m-1$
peripheral nodes of each copy being connected to the root of the
original $M_{1}$. In this way, we obtain $M_{2}$, the hub and
peripheral nodes of which are the hub of the original $M_{1}$ and
the $(m-1)^2$ peripheral nodes in the $m-1$ duplicates of $M_{1}$,
respectively. Supposing one has $M_{g-1}$, the next generation network
$M_{g}$ can be obtained by adding $m-1$ copies of $M_{g-1}$ to the
primal $M_{g-1}$, with all peripheral nodes of the replicas being
linked to the hub of the original $M_{g-1}$ unit. The hub of the
original $M_{g-1}$ and the peripheral nodes of the $m-1$ copies of
$M_{g-1}$ form the hub node and peripheral nodes of $M_{g}$,
respectively. Repeating indefinitely the two steps of replication
and connection, one obtains the modular scale-free networks.
Figure~\ref{network} illustrates a network $M_4$ for the particular
case of $m=5$.

Many interesting quantities and properties of the model can be
determined explicitly~\cite{RaBa03,No03}. In $M_{g}$, the network
size (number of nodes), denoted by $N_g$, is $N_g=m^{g}$. All these nodes can be
classified into four distinct sets~\cite{No03,NoRi04a}: the peripheral
node set $\mathbb{P}$, the locally peripheral node set $\mathbb{P}_z$
($1\leq z < g$), the set $\mathbb{H}$ containing only the hub node of
$M_{g}$, and the local hub set $\mathbb{H}_z$ ($1\leq z < g$). The number of nodes in each of these four sets is
\begin{equation}\label{Car_K_p}
|\mathbb{P}| = (m-1)^{g},
\end{equation}
\begin{equation}\label{Car_K_lp}
|\mathbb{P}_z| = (m-1)^{z}m^{g-(z+1)},
\end{equation}
\begin{equation}\label{Car_K_h}
|\mathbb{H}| = 1,
\end{equation}
and
\begin{equation}\label{Car_K_lh}
|\mathbb{H}_z| = (m-1) m^{g-(z+1)},
\end{equation}
respectively. For $M_{g}$, all nodes in a set have the same degree.
It has been obtained exactly that the degree for a node in sets
$\mathbb{H}$, $\mathbb{H}_z$, $\mathbb{P}$, and $\mathbb{P}_z$ is, respectively,
\begin{equation}\label{K_h}
K_h(g) = \sum_{g_i=1}^{g}(m-1)^{g_i}=\frac{m-1}{m-2}[(m-1)^g-1]\,,
\end{equation}
\begin{equation}\label{K_lh}
K_{h,z}(g) = \sum_{g_i=1}^{z}(m-1)^{g_i}=\frac{m-1}{m-2}
[(m-1)^z-1]\,,
\end{equation}
\begin{equation}\label{K_p}
K_p(g)=g+m-2\,,
\end{equation}
and
\begin{equation}\label{K_lp}
K_{p,z}(g)=z+m-2\,.
\end{equation}
In addition, it is easy to obtain that the average degree of all
nodes is approximately equal to a constant $2(m-1)(3m-2)/m$ in the
limit of infinite $g$, showing that the networks are sparse.

The model under consideration is in fact an extension of the one
proposed in Ref.~\cite{BaRaVi01} and studied in much detail
in Refs.~\cite{IgYa05,ZhLiGaZhGu09,AgBuMa10}. It presents some typical features
observed in a variety of real-world systems~\cite{RaBa03,No03}. Its
degree distribution follows a power-law scaling $P(k) \sim
k^{-\gamma}$ with a general exponent $\gamma=1+\ln m/\ln (m-1)$
belonging to the interval $(2,2.585]$. Its average clustering
coefficient tends to a large constant dependent on $m$; and its
average distance grows logarithmically with the network order~\cite{ZhLi10}, both of which show that the model is small world~\cite{WaSt98}. In addition, the betweenness
distribution $P(b)$ of nodes also obeys a power-law behavior $P(b)
\sim b^{-2}$ with the exponent independent of the parameter $m$.
In particular, the whole class of networks shows a remarkable
modular structure. These peculiar structural properties make the
networks unique within the category of complex networks. It is thus
interesting to address dynamical processes happening on them. The main purpose of this work is to study random walks on this network family
with multiple traps located on some special nodes.

\section{Formulation of random walks with multiple traps on a network}

In this section, we formulate the problem of random walks on the
network $M$ with multiple traps, which is a
discrete-time random walk of a particle in the presence of several perfect
traps placed on certain nodes. At each time step, the particle jumps
with equal probability from its current location to one of its
nearest neighbors. If the particle meets one of the traps, then it
is absorbed. At last, the particle will be inevitably
absorbed by the traps, regardless of its starting
position~\cite{KeSn76,AlFi99}.

It is well known that an arbitrary network can be completely
represented by its adjacency matrix. For $M$, its adjacency matrix
$\textbf{A}$ is a matrix consisting of entries 0 or 1, with an
order $N \times N$ ($N$ is the number of nodes in $M$). The $(i,j)$ element $a_{ij}$ of
$\textbf{A}$ is defined as follows:
$a_{ij}=1$ if $i$ and $j$ are neighbors and
$a_{ij}=0$ otherwise. Then the degree $d_{i}$ of node $i$ is
given by $d_{i}=\sum_{j=1}^{N}a_{ij}$, the diagonal degree
matrix $\textbf{Z}$ associated with $M$ is $\textbf{Z}={\rm
diag} (d_1, d_2,\ldots, d_i, \ldots, d_{N})$, and the
corresponding normalized Laplacian matrix of $M$ is defined to be
$\textbf{L}=\textbf{I}-\textbf{Z}^{-1}\,\textbf{A}$, where
$\textbf{I}$ is the identity matrix with order $N \times N$.

We use $\Gamma$ to denote the set of traps and $|\Gamma|$ to represent the number of traps.
We are concerned with the expected time the particle spends,
starting from a source node, before it falls on one of
the traps for the first time. Let $T_i$ be the expected time,
frequently called first-passage time (FPT) or trapping time, for a
particle first arriving at any one of the traps, given that it
starts from node $i$. It is clear that for any node $i \in
\Gamma$, we have $T_i=0$. The set of this important quantity
satisfies the relation
\begin{equation}\label{MFPT1}
 T_i=\sum_j w_{ij}\,T_j+1\,,
\end{equation}
where $i \bar{\in} \Gamma$ and $w_{ij}$ is transition probability
for the particle of going from node $i$ to node $j$. According to
the definition of the random-walk problem, it is not difficult to
know that $w_{ij}=a_{ij}/{d_i}$, which is exactly the
$(i,j)$ element of the matrix $\textbf{Z}^{-1}\,\textbf{A}$.

In order to facilitate the description, we distinguish all nodes in
$M$ by assigning each of them a unique number. We label
consecutively all nodes, excluding those in $\Gamma$, from 1 to
$N-|\Gamma|$ and trap nodes are
numbered from $N-|\Gamma|+1$ to $N$. Then
Eq.~(\ref{MFPT1}) can be rewritten in matrix form as
\begin{equation}\label{MFPT2}
\textbf{T}'=\textbf{W}'\,\textbf{T}'+\textbf{e},
\end{equation}
where
$\textbf{T}'=\left[T_1,T_2,\ldots,T_{N-|\Gamma|}\right]^\top$
(the superscript $\top$ of the vector represents transpose) is an
$(N-|\Gamma|)$-dimensional vector, $\textbf{e}$ is the
$(N-|\Gamma|)$-dimensional unit vector $(1,1,\ldots,1)^\top$, and
$\textbf{W}'$ is the transition matrix corresponding to the trapping problem. Equation~(\ref{MFPT2}) can
be further recast as
\begin{equation}\label{MFPT3}
 \textbf{T}'=[\textbf{L}']^{-1}\,\textbf{e},
\end{equation}
where
\begin{equation}\label{MFPT4}
 \textbf{L}'=\textbf{I}'-\textbf{W}'
\end{equation}
with $\textbf{I}'$ being the $(N-|\Gamma|)\times
(N-|\Gamma|)$ identity matrix.

It should be mentioned that the considered random walk is in fact a
Markov process, and Eq.~(\ref{MFPT4}) is the fundamental matrix of
the Markov chain representing such an unbiased random walk. We also
note that the matrix $\textbf{I}'-\textbf{W}'$ on the right-hand
side of Eq.~(\ref{MFPT4}) is actually a submatrix of the
normalized discrete Laplacian matrix $\textbf{L}$ of $M$, which
is obtained from $\textbf{L}$ by suppressing the last $|\Gamma|$
rows and columns that correspond to the trap nodes.

Equation~(\ref{MFPT3}) shows that trapping time $T_i$ can be expressed in
terms of the entries $l_{ij}^{-1}$ of inverse matrix of
$\textbf{L}'$ (i.e., a submatrix of
$\textbf{L}$). Concretely, $T_i$ is provided by
\begin{equation}\label{MFPT5}
T_i=\sum_{j=1}^{N-|\Gamma|}l_{ij}^{-1}\,,
\end{equation}
which accounts for the Markov chain representing the random walk:
The entry $l_{ij}^{-1}$ of the fundamental matrix
$[\textbf{L}']^{-1}$ for the Markov process represents the expected
number of times the particle visits node $j$ in the case that it
starts off from node $i$ (see Ref.~\cite{BaKl98} for a single trap).
Then the MFPT $\langle T \rangle$, which is defined as the average of
$T_i$ over all initial nodes distributed uniformly over nodes in
$M$ including the traps~\cite{note}, is given by
\begin{eqnarray}\label{MFPT6}
\langle T \rangle
&=&\frac{1}{N}\sum_{i=1}^{N}T_i=\frac{1}{N}\sum_{i=1}^{N-|\Gamma|}T_i\nonumber \\
&=& \frac{1}{N}\sum_{i=1}^{N-|\Gamma|}\sum_{j=1}^{N-|\Gamma|}{l_{ij}^{-1}}\,.
\end{eqnarray}
Thus, on the basis of the definition of the unbiased random walks, we
have derived the numerical yet exact solution to the MFPT $\langle T
\rangle$ for random walks on any network with multiple
traps, independently of the number and location of the traps. We note that our derivation is a reformulation of the backward equation satisfied by the MFPT and that Eq.~(\ref{MFPT6}) can also be found in the literature in several equivalent forms~\cite{KeSn76,AlFi99}.

Equation~(\ref{MFPT6}) is very important, since it reduces the
problem of calculating the MFPT $\langle T \rangle$ to computing the
sum of the elements of the matrix $[\textbf{L}']^{-1}$ and can be used to check the results for $\langle T \rangle$ derived  by  other methods, at least for networks with a small number of nodes.
However, it is notable that although the above computational method,
process, and result are applicable to the trapping issue on all
networks, the derivation of Eq.~(\ref{MFPT6})
requires inverting the matrix $\textbf{L}'$ with an order
$(N-|\Gamma|) \times (N-|\Gamma|)$. Since the computation of
inverting the matrix $\textbf{L}'$ puts heavy demands on time and
memory for large networks, by using Eq.~(\ref{MFPT6}) we can directly
compute $\langle T \rangle$ only for networks with small size. In
particular, by applying the approach of inverting the matrix, it appears
very difficult, even impossible, to get the exact dominating
scaling of $\langle T \rangle$ characterizing the efficiency of
the trapping problem. Therefore, it is of significant practical
importance to seek an alternative method of computing $\langle T \rangle$ even for specific networks,
which is able to reduce the computational effort of the method
of inverting the matrix.

\section{Scalings of the MFPT for random walks on modular networks with multiple traps}

Here we study two particular trapping problems defined in the modular scale-free networks $M_g$. We first address the case that traps are located at all peripheral nodes; then we consider the case that traps are fixed on those nodes farthest from the main hub. We will show that the special recursive construction of the modular scale-free networks and the particular
selections made for the trap locations allow for an analytical
treatment of the MFPT to the traps.

\subsection{Determination of intermediate variables}

Prior to studying the MFPT to the traps, we first define some intermediate variables and determine their values. We denote by $T_g^P$ the FPT for a walker starting from
an arbitrary peripheral node of $M_g$ to visit the hub for the first
time and by $T_g^H$ the FPT spent by a particle initially
located at the hub to first visit any peripheral node. In Appendix~\ref{App01}, we give detailed derivations for $T_g^H$ and $T_g^P$, which read
\begin{equation}\label{MTTa08}
T_{g}^H=\left(3-\frac{5m-2}{m^2}\right)\left(\frac{m}{m-1}\right)^{g}-1
\end{equation}
and
\begin{equation}\label{MTTa09}
T_{g}^{P}=\left(3m-8+\frac{7m-2}{m^{2}}\right)\left(\frac{m}{m-1}\right)^{g}-2m+3,
\end{equation}
respectively.

Equations~(\ref{MTTa08}) and~(\ref{MTTa09}) are very useful for the following derivation of the
exact formula for the MFPT to the targets. We note that Eqs.~(\ref{MTTa08}) and~(\ref{MTTa09}) have also been derived in Ref.~\cite{ZhLiGoZhGuLi09} by using the technique of generating functions~\cite{Wi94}, but the approach used here is different from and relatively easier than the previous one.

\subsection{Exact solution to the MFPT for random walks with traps located at peripheral nodes}

After obtaining the intermediate quantities, we are now in a position to consider random walks on networks $M_g$ with all the $(m-1)^g$ peripheral nodes being occupied by traps. Our goal in this case is to determine the MFPT denoted by $\langle T \rangle_g$, which is the average of the FPT for a walker originating from a node in $M_g$ to first visit any target over all starting points including the traps. In order to find $\langle T \rangle_g$, we introduce another quantity $\langle H \rangle_g$ defined as the FPTs of all nodes to the hub. From the structure of the networks, we can easily establish the following recursive relations for $\langle T \rangle_g$ and $\langle H \rangle_g$:
\begin{equation}\label{MTTb01}
\langle T \rangle_g=\frac{1}{m}\left (\langle H \rangle_{g-1}+ T_{g}^{H}\right)+\frac{m-1}{m} \langle T \rangle_{g-1}
\end{equation}
and
\begin{equation}\label{MTTb02}
\langle H \rangle_g=\frac{1}{m}\langle H \rangle_{g-1} + \frac{m-1}{m}\left(\langle T \rangle_{g-1} + T_{g}^{P} \right)\,.
\end{equation}

Equations~(\ref{MTTb01}) and~(\ref{MTTb02}) can be rewritten as
\begin{equation}\label{MTTb03}
m\langle T \rangle_g-(m-1)\langle T \rangle_{g-1}- T_{g}^{H}=\langle H \rangle_{g-1}
\end{equation}
and
\begin{equation}\label{MTTb04}
m\langle H \rangle_g-\langle H \rangle_{g-1} =(m-1)\left(\langle T \rangle_{g-1} + T_{g}^{P} \right)\,.
\end{equation}
From Eq.~(\ref{MTTb03}), we further have
\begin{equation}\label{MTTb05}
m\langle T \rangle_{g+1}-(m-1)\langle T \rangle_{g}- T_{g+1}^{H}=\langle H \rangle_{g}\,,
\end{equation}
which, together with Eqs.~(\ref{MTTb03}) and~(\ref{MTTb04}), yields
\begin{eqnarray}\label{MTTb06}
&\quad& m\left[m\langle T \rangle_{g+1}-(m-1)\langle T \rangle_{g} - T_{g+1}^{H} \right]-\nonumber \\ &\quad&\left[ m\langle T \rangle_{g}-(m-1)\langle T \rangle_{g-1} - T_{g}^{H} \right] \nonumber \\&=& m\langle H \rangle_{g} - \langle H \rangle_{g-1}\nonumber \\
 &=&(m-1)\left(\langle T \rangle_{g-1}+ T_{g}^{P}\right)\,,
\end{eqnarray}
that is,
\begin{equation}\label{MTTb07}
\langle T \rangle_{g+1}-\langle T \rangle_{g}=\frac{1}{m^2}\left[mT_{g+1}^{H} - T_{g}^{H} + (m-1)T_{g}^{P} \right].
\end{equation}

Substituting Eqs.~(\ref{MTTa08}) and~(\ref{MTTa09}) into
Eq.~(\ref{MTTb07}) and considering the initial condition $\langle T \rangle_{2}=m+1-2/m$, we can solve inductively Eq.~(\ref{MTTb07}) to obtain the following rigorous expression:
\begin{eqnarray}\label{MTTb08}
\langle T \rangle_{g}&=&\frac{(m-1)(3m-2)(m^2-2m+2)}{m^3} \left(\frac{m}{m-1}\right)^g \nonumber \\ &\quad&
- \frac{2(m-1)^2}{m^2}g-3m+10-\frac{12}{m}+ \frac{4}{m^2}\,.
\end{eqnarray}
Plugging this result for $\langle T \rangle_{g}$ into Eq.~(\ref{MTTb02}) and using the initial condition $\langle H \rangle_{2}=m/(m-1)$, Eq.~(\ref{MTTb02}) is solved to yield
\begin{eqnarray}\label{MTTb09}
\langle H \rangle_g &=&\frac{2(3m-2)(m-1)^3}{m^3}\left(\frac{m}{m-1} \right)^g \nonumber \\
&\quad&- \frac{2(m-1)^2}{m^2}g-\frac{m-1}{m^2}\left(5m^2- 10m+4\right)\, .
\end{eqnarray}

\begin{figure*}
\begin{center}
\includegraphics[width=0.85\linewidth, trim=50 20 0 15]{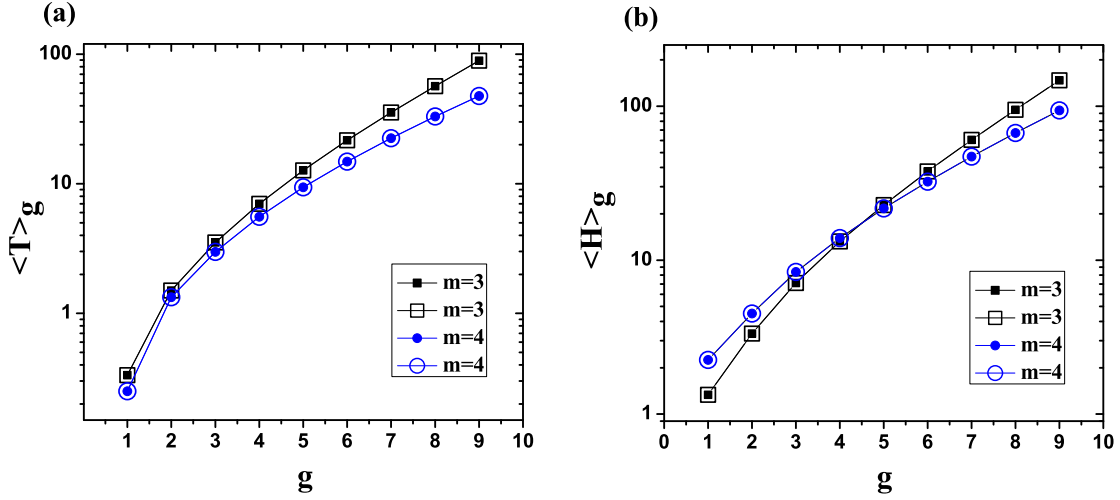}
\end{center}
\caption[kurzform]{\label{Time01} (Color online) Mean first-passage
times $\langle T \rangle_g$ and $\langle H \rangle_g$ as functions of the iteration $g$ on a
log-log scale for the two cases of $m=3$ and $4$. The open
symbols represent the numerical results obtained by direct
calculation from Eq.~(\ref{MFPT6}); the solid symbols
correspond to the rigorous values given by Eq.~(\ref{MTTb08}) or~(\ref{MTTb09}).}
\end{figure*}

To confirm our analytic formulas, we have compared them with the
numerical values from the method of inverting the matrix provided by
Eq.~(\ref{MFPT6}); see Fig.~\ref{Time01}. For various values of $m$ and $g$, the results
for $\langle T \rangle_{g}$ and $\langle H \rangle_{g}$ obtained separately from Eqs.~(\ref{MTTb08}) and~(\ref{MTTb09}) are in
complete agreement with those from Eq.~(\ref{MFPT6}). This agreement serves as a mutual test of our numerical solution and analytical formulas, providing an important
evidence of the validity of Eqs.~(\ref{MFPT6}),~(\ref{MTTb08}) and~(\ref{MTTb09}).

We proceed to represent $\langle T \rangle_{g}$ and $\langle H \rangle_{g}$ as functions of
network size $N_g$ to obtain their dependence on $N_g$. From $N_g=m^g$ we have $g=\ln N_g/ \ln m$ and $m^g/(m-1)^g=(N_g)^{1-\ln(m-1)/\ln m}$, which
enables us to recast Eqs.~(\ref{MTTb08}) and~(\ref{MTTb09}) in terms of $N_g$ as
\begin{eqnarray}\label{MTTb10}
\langle T \rangle_{g}&=&\frac{(m-1)(3m-2)(m^2-2m+2)}{m^3} (N_g)^{1-\ln(m-1)/\ln m} \nonumber \\ &\quad&
- \frac{2(m-1)^2}{m^2}\frac{\ln N_g}{\ln m}-3m+10-\frac{12}{m}+ \frac{4}{m^2}
\end{eqnarray}
and
\begin{eqnarray}\label{MTTb11x}
\langle H \rangle_g &=&\frac{2(3m-2)(m-1)^3}{m^3}(N_g)^{1-\ln(m-1)/\ln m} \nonumber \\
&\quad&- \frac{2(m-1)^2}{m^2}\frac{\ln N_g}{\ln m}-\frac{m-1}{m^2}\left(5m^2- 10m+4\right)\,.\nonumber \\
\end{eqnarray}

Equations~(\ref{MTTb10}) and~(\ref{MTTb11x}) imply that in the limit of large network
size (i.e., $N_g \rightarrow \infty$), both $\langle T
\rangle_g$ and  $\langle H \rangle_g$ grow asymptotically as power-law functions of
network size $N_g$ with the same exponent $\eta(m)=1-\ln(m-1)/\ln m$:
\begin{equation}\label{MTTb11}
\langle T \rangle_g \sim (N_g)^{\eta(m)}=(N_g)^{1-\ln(m-1)/\ln m},
\end{equation}
and
\begin{equation}\label{MTTb12}
\langle H \rangle_g \sim (N_g)^{\eta(m)}=(N_g)^{1-\ln(m-1)/\ln m}\,.
\end{equation}
Obviously, the exponent $\eta(m)$ is smaller than 1, showing that both $\langle
T \rangle_{g}$ and $\langle
H \rangle_{g}$ scale sublinearly with the network size.

We note that the scaling in Eq.~(\ref{MTTb11}) has been previously derived in Ref.~\cite{ZhLiGoZhGuLi09} by using the theory of generating functions, the computation process of which is a little complex. Equation~(\ref{MTTb11}) shows that when the hub node is considered an immobile trap, the trapping efficiency is high (even the highest among all networks~\cite{TeBeVo09,ZhAlHoZhCh11}), which can be elaborated as follows. In $M_g$ the average distance between the hub and other nodes is only half of the average distance between all pairs on nodes~\cite{ZhLi10}, suggesting that the hub node is spatially closer than any other node. In contrast, the degree of the hub node is the highest, which is why the MFPT to the hub is very low.

\subsection{Behavior of the MFPT for random walks with traps fixed at farthest nodes}

We now focus on the trapping problem with traps being placed on the nodes farthest from the main hub, which are expected be more difficult to visit compared with the peripheral nodes~\cite{TeBeVo11}.

\subsubsection{Related definitions and quantities}

In $M_g$ the maximum value of the distance from the main hub to other nodes is $g$. We let $F_g$ denote the set of those nodes in $M_g$ at a  distance $g$ from the main hub of $M_g$, hereafter called the farthest nodes of $M_g$, and $|F_g|$ denote the cardinality (number of elements in a set) of $F_g$. By construction, $M_g$ is composed of a primal $M_{g-1}$ and $m-1$ copies of $M_{g-1}$, denoted separately by $M_{g-1}^{(x)}$ ($x=1,2,\ldots, m-1$). For $M_1$, its farthest nodes are exactly its $m-1$ peripheral nodes; for $M_2$, its farthest nodes correspond to the hub nodes of all $M_{1}^{(x)}$. Proceeding analogously, for $g \geq 3$ the farthest nodes of $M_g$ must belong to all subgraphs $M_{g-1}^{(x)}$, and the farthest nodes of the primal central subgraphs (i.e., $M_{g-2}$) forming $M_{g-1}^{(x)}$ constitute $F_g$. Thus, we have
\begin{equation}\label{MTTc01}
|F_g|=(m-1)|F_{g-2}|\,.
\end{equation}
Considering $|F_1|=m-1$ and $|F_2|=m-1$, the recursive relation can be solved to obtain
\begin{equation}\label{MTTc02}
|F_g|=
\begin{cases}
(m-1)^{(g+1)/2}, &g \quad {\rm is \quad odd},\\
(m-1)^{g/2}, &g \quad {\rm is \quad even}.
\end{cases}
\end{equation}

Next we concentrate on the MFPT from the hub to the farthest nodes in $M_g$, which will be denoted by $\langle T\rangle _g^{H}$ henceforth since, as will be shown, it has the same scaling as that of the average of MFPTs to the farthest nodes $F_g$, taken over all starting points. For a convenient description of the computation for the MFPT to the farthest nodes, we introduce more variables. For $M_g$, let $H_g$ and $R_g$ express the sets of the main hub and peripheral nodes, respectively. In addition, for those nodes of $M_g$ that belong to $M_{g-1}^{(x)}$, we can further classify them in the following way. Let $H_{g-n}$ ($n=1,2,\ldots, g-1$) denote the set of those local hubs that are directly connected to $g-n$ classes of local peripheral nodes in $\mathbb{P}_z$ and $R_{g-n}$ ($n=1,2,\ldots, g-1$) stand for the set of the local peripheral nodes whose neighbors are $g-n$ different local hubs belonging to $\mathbb{H}_z$. It is easy to verify that the respective degrees of nodes in $R_{g-n}$ and $H_{g-n}$ are $K_{g-n}^{R}=m-2+g-n$ and $K_{g-n}^{H}=\sum_{i=1}^{g-n}(m-1)^{i}$, respectively.

\subsubsection{Exact solution to the MFPT from the hub to farthest nodes}

According to the structure of $M_g$, for a walker starting from the main hub, in order to reach the farthest nodes, it should follow the path $H_g \rightarrow R_g \rightarrow H_{g-1}\rightarrow R_{g-2}\rightarrow H_{g-3} \cdots \rightarrow R_{g-(n-2)}\rightarrow H_{g-(n-1)} \rightarrow R_{g-n} \rightarrow H_{g-(n+1)} \rightarrow R_{g-(n+2)}\cdots R_1 ({\rm or } \, H_1)$. Then it is natural to define the following quantities. Let $R_g(n)$ and $H_g(n)$ represent, respectively, the FPT from a node in $R_{g-n}$ to any of its neighboring nodes in $H_{g-(n+1)}$ and the FPT from a node in $H_{g-n}$ to any of its neighbors belonging to $R_{g-(n+1)}$. In Appendix~\ref{App02}, we report the derivation for $R_g(n)$ and $H_g(n)$, the exact expressions for which are
\begin{eqnarray}\label{MTTc07}
R_{g}(n)&=&(m-1)^{n/2+1} \left[ \frac{3m-2}{ m } \left(\frac{m}{m-1}\right)^g - 2 \right]\nonumber \\  &\quad& - \frac{3m - 2}{m-1}  \left( \frac{m}{m-1} \right)^{g-n-3} + 1
\end{eqnarray}
and
\begin{eqnarray}\label{MTTc08}
H_{g}(n)&=&(m-1)^{(n + 3)/2} \left[  \frac{3m - 2 }{m}\left(\frac{m}{m-1}\right)^{g} - 2 \right] \nonumber \\ &\quad& -(3 m-2) \left(\frac{m}{m-1}\right)^{g-n-3} + 2m-3   \,,
\end{eqnarray}
respectively.

Using the obtained intermediate quantities, we can derive an exact formula for $\langle T\rangle _g^{H}$. We distinguish two cases: (i) $g$ is odd and (ii) $g$ is even. For odd $g$ we have
\begin{eqnarray}\label{MTTc09}
\langle T \rangle_g^{H} &=&T_{g}^{H}+\sum_{i=0}^{(g-1)/2-1}R_{g}(2i)+\sum_{i=0}^{(g-1)/2-2}H_{g}(2i+1)\nonumber \\ &\quad&+(m-1)R_{g}(g-3)+m
\end{eqnarray}
By plugging Eqs.~(\ref{MTTc07}) and~(\ref{MTTc08}) into Eq.~(\ref{MTTc09}) and doing some algebra, we find a closed-form solution to $\langle T \rangle_g^{H}$ given by
{\scriptsize\
\begin{eqnarray}\label{MTTc10}
\langle T \rangle_g^{H} &=&\frac{3m-2}{m-2} \left(\frac{m}{m-1}\right)^g (m-1)^{(g+1)/2}-\frac{2 m}{m-2}(m-1)^{(g+1)/2}\nonumber \\ &\quad&-\frac{(m-1)(3m-2)\left(m^2-2m+3\right)}{(m-2)(2m-1)}\left(\frac{m}{m-1}\right)^g+\nonumber \\
&\quad& (m-1)g + \frac{(m-1)\left(3m^3-9m^2+14m-4\right)}{(m-2)(2m-1)}.
\end{eqnarray}}

\begin{figure}
\begin{center}
\includegraphics[width=1.05\linewidth, trim=15 40 20 25]{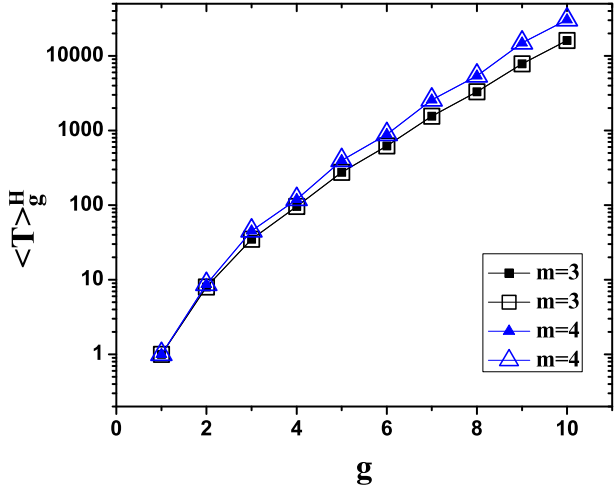}
\end{center}
\caption[kurzform]{\label{Time02} (Color online) Mean first-passage
time $\langle T \rangle_g^{H}$ as a function of generation $g$ on a
log-log scale for two special cases of $m=3$ and $4$. The open
symbols indicate the numerical results obtained by direct
calculation from Eq.~(\ref{MFPT5}); the solid symbols
display the analytical values provided by Eqs.~(\ref{MTTc10}) and~(\ref{MTTc12}).}
\end{figure}

When $g$ is even, it is not difficult to reach the following expression:
\begin{eqnarray}\label{MTTc11}
\langle T\rangle_g^{H} &=& T_g^{H}+\sum_{i=0}^{g/2-2}R_{g}(2i)+\sum_{i=0}^{g/2-2}H_{g}(2i+1)\nonumber \\ &\quad&+H_{g}(g-3)+m\,.
\end{eqnarray}
Inserting Eqs.~(\ref{MTTc07}) and~(\ref{MTTc08}) into Eq.~(\ref{MTTc11}), after some algebra, the explicit expression for $\langle T\rangle_g^{H}$ is obtained, which reads
{\scriptsize\
\begin{eqnarray}\label{MTTc12}
\langle T \rangle_g^{H} &=&\frac{2(3 m-2)}{(m-2)m}\left( \frac{m}{m-1} \right) ^g (m-1)^{g/2+1} -\frac{4}{m-2}(m-1)^{g/2+1}\nonumber \\
&\quad&-\frac{ (m-1)(3 m-2) \left(m^2-2 m+3\right)}{(m-2) (2 m-1)}\left(\frac{m}{m-1}\right)^g +\nonumber \\
&\quad& (m-1)g + \frac{3 m^4-11 m^3+19 m^2-14 m+4}{(m-2) (2 m-1)}.
\end{eqnarray}}
To check the validity of Eqs.~(\ref{MTTc10}) and~(\ref{MTTc12}), we also compute $\langle T \rangle_g^{H}$ numerically by using the approach of inverting the related matrix; see Eq.~(\ref{MFPT5}). The results obtained by analytical and numerical methods completely agree with each other. The comparison is shown in Fig.~\ref{Time02}. Equation~(\ref{MTTc10}), together with Eq.~(\ref{MTTc12}), indicates that
for large networks, i.e., $N_g\rightarrow \infty$,
\begin{equation}\label{MTTc13}
\langle T \rangle_g^{H}  \sim (N_g)^{\theta(m)}=(N_g)^{1-\ln (m-1)/(2\ln m)},
\end{equation}
with the exponent $\theta(m)=1-\ln (m-1)/(2\ln m)$ smaller than 1.

Thus far we have found the rigorous formula for  the MFPT $\langle T \rangle_g^{H}$ to farthest nodes in $M_g$ and its dependence on network size $N_g$. We stress that the analytical  computation for the MFPT $\langle T \rangle_g$ to the farthest nodes that average all starting points in $M_g$ is rather lengthy and awkward. However, it is easy to infer that when $g$ is large enough, the dominant term of $\langle T \rangle_g$ also increases as a power-law function of network size $N_g$ with an exponent identical to that of $\langle T \rangle_g^{H}$, which can be understood from the following heuristic explanation. Note that $M_g$ consists of $m$ subgraphs, which are copies of  $M_{g-1}$. For those nodes in the central subgraph, their MFPT to the farthest nodes is equal to $\langle T \rangle_g^{H}+\langle H \rangle_{g-1}$, the dominant term of which is $\langle T \rangle_g^{H}$; for nodes in each of the $m-1$ fringe subgraphs $M_{g-1}^{(x)}$ ($x=1,2,\ldots, m-1$), their MFPT to the farthest nodes is identical but smaller than $\langle T \rangle_g^{H}$. Hence, for all nodes in $M_g$, the dominating term of the MFPT $\langle T \rangle_g$ is proportional to $(N_g)^{\theta(m)}$ but its prefactor may be different from that of $\langle T \rangle_g^{H}$.

\subsection{Result analysis}

Equations~(\ref{MTTb11}) and~(\ref{MTTc13}) show that when traps are positioned at several particular nodes, the MFPTs to the target node are very small, which scale sublinearly with the network order. When either peripheral nodes or farthest nodes are occupied by traps, the characteristic exponent $\eta(m)$ or $\theta(m)$ is a decreasing function of $m$: When the parameter $m$
increases from 3 to $\infty$, both $\eta(m)$ and $\theta(m)$ drop
and are close to zero. Therefore, the efficiency of the random-walk
process is reliant on $m$: The larger the parameter $m$, the more
efficient the random-walk process. The fact that both trapping processes are very efficient  demonstrates that the modular
scale-free networks being studied exhibit an efficient configuration for random walks
with traps positioned at certain given nodes.

In contrast, for each given parameter $m$, $\eta(m)$ is smaller than $\theta(m)$, which implies that when traps are located at peripheral nodes, the trapping efficiency is higher than that of the case when traps are placed on farthest nodes. Thus, the two trapping processes defined on the networks under consideration display rich behavior in the context of MFPTs to the traps. The difference between $\eta(m)$ and $\theta(m)$ shows that the number and location of traps sensitively affect the behavior of random walks on the modular scale-free networks.

Actually, the intrinsic structure of the modular scale-free networks is
responsible for the high efficiency of random walks performing on
them with certain nodes being occupied by traps. In these
networks, there are many small highly integrated clusters, which
group into a few larger but less compact modules linked by local hub
nodes; see Fig.~\ref{network}. These relatively large modules
combine to form even larger and fewer groups, which are further
joined to shape a fine modular and scale-free architecture, a
topology that accounts for the fast diffusion phenomenon in $M_g$.

In the case that traps are placed on peripheral nodes, when a particle originates from a node in a duplicate $M_{g-1}$ (an
element of $M_g$), it will either be directly trapped by one of the
traps or jump to local hub nodes in a few steps. These local hubs
play a bridge role linking different small modules together. After
arriving at local hub nodes, the particle can be easily trapped in a
short time. In contrast, if the particle starts off from a
node in the original $M_{g-1}$ (the central part of $M_g$), it will
easily visit local hub nodes or the hub first, through which it can
find the way to one of the traps quickly. Thus, the particle can
drop into the traps very fast wherever it starts to jump, which can
be understood from the above heuristic argument based on the
inherent structure of the considered networks.

When the traps are fixed on the farthest nodes, to find a garget, the walker must first visit the local hubs and local peripheral nodes of a larger and sparser cluster, starting from which it continues to arrive at the local hubs and local peripheral nodes of smaller and denser groups. From Eqs.~(\ref{MTTc07}) and~(\ref{MTTc08}) we know that the expected time between local hub nodes and local peripheral nodes in inner subgraphs rely on their size or deepness (i.e., $g-n$): The smaller the value of $g-n$, the smaller the size of inner subgraphs, and the higher the expected time. This can account for the main reason the farthest nodes are more difficult to reach than the peripheral nodes.

\section{Conclusions}

We have studied the random-walk dynamics on a family of modular scale-free networks with multiple traps, which exhibit remarkable characteristics observed for various real-life networks, such as social and biological networks. We first deduced a general formula for the MFPT to the traps in a generic network, which is expressed in terms of several elements of a matrix associated with the trapping problem. Then we studied the MFPT for two trapping issues on the studied networks with two different arrangements of targets. In the first case, peripheral nodes are treated as traps; in the second case, farthest nodes work as traps.

For the two trapping problems, we studied both numerically and analytically the MFPT to traps, the results of which are compatible with each other. Our results show that in both cases, the MFPT varies as a power-law function of network size with the exponent depending on the parameter $m$, which is lower than 1 in the full range of $m$. Thus, the studied networks display an efficient architecture in favor of diffusion. Moreover, we demonstrated that, compared with the second case, the diffusion is faster in the first case, which indicates that the transport efficiency relies on the number and location of the absorbing nodes. We also showed that the modular topology, together with the scale-free behavior, is responsible for the quick diffusion processes, as well as the scaling difference of the MFPT for the two trappings running on the networks addressed. We expect that our work can provide insight into designing networks with a structure in favor of diffusion. Finally, it should be mentioned that the method developed here applies only to very
specific sets of traps and is hard to generalize to other sets of traps.

\subsection*{Acknowledgment}

This research was supported by the National Natural Science Foundation of China under Grant No. 61074119.

\appendix

\section{Derivation of $T_g^P$ and $T_g^H$ \label{App01}}

According to the particular structure of the networks, for any $g >1$, the two
quantities $T_g^P$ and $T_g^H$ obey the following recursion relations:
\begin{eqnarray}\label{MTTa01}
T_{g}^{P}&=&\frac{1}{(m-2)+g}\bigg[1+(m-2)\left(1+T_{g}^{P}\right)+\nonumber \\ &\quad& \sum_{i=1}^{g-1}{\left(1+T_{i}^{H}+T_{g}^{P}\right)} \bigg]
\end{eqnarray}
and
\begin{eqnarray}\label{MTTa02}
T_{g}^{H}&=&\frac{1}{\sum_{i=1}^{g}{(m-1)^i}} \bigg [(m-1)^g+ \nonumber \\ &\quad& \sum_{i=1}^{g-1}(m-1)^i \left(1+T_{i}^{P}+T_{g}^{H}\right)\bigg ].
\end{eqnarray}

The three terms on the right-hand side (rhs) of Eq.~(\ref{MTTa01})
can be explained as follows. The first term is based on the fact that the walker takes only one time step to first reach the hub. The second term describes the process by which the
particle first jumps to one of its $m-2$ neighbors belonging to
$\mathbb{P}$ in one time step and then takes $T_{g}^{P}$ more steps to
first get to the target node. The last term accounts for
the fact that the walker first makes a
jump to a local hub node belonging to $\mathbb{H}_z$, then takes
$T_{i}^{H}$ time steps, starting off from the local hub, to reach any node
in $\mathbb{P}$, and continues to jump $T_{g}^{P}$ more steps to
reach the target node for the first time.

Analogously, the two terms on the rhs of Eq.~(\ref{MTTa02}) are based on the following two processes. The first term describes the fact that the walker,
starting from the hub, requires only one time step to hit a
peripheral node. The second term explains such a process that the walker,
starting off from the hub, first jumps to a local peripheral
node belonging to $\mathbb{P}_z$ in one time step, then makes $T_{i}^{P}$ jumps to the hub, and
proceeds to any node in $\mathbb{P}$, taking $T_{g}^{H}$ more time steps.

After merging similar items, Eqs.~(\ref{MTTa01}) and~(\ref{MTTa02}) can be rewritten as
\begin{equation}\label{MTTa03}
T_{g}^{P}=(m-2)+g+\sum_{i=1}^{g-1}{T_{i}^{H}}
\end{equation}
and
\begin{equation}\label{MTTa04}
T_{g}^{H}=\frac{1}{(m-1)^g}\left[ \sum_{i=1}^{g}{(m-1)^i}+\sum_{i=1}^{g-1}{(m-1)^iT_{i}^{P}} \right],
\end{equation}
respectively.
Equations~(\ref{MTTa03}) and~(\ref{MTTa04}) lead to
\begin{equation}\label{MTTa05}
T_{g+1}^{P}-T_{g}^{P}=1+T_{g}^{H}
\end{equation}
and
\begin{equation}\label{MTTa06}
T_{g+1}^{H}-\frac{1}{m-1}T_{g}^{H}=1+\frac{1}{m-1}T_{g}^{P}\,.
\end{equation}
According to Eq.~(\ref{MTTa06}), we obtain
\begin{eqnarray}\label{MTTa07}
&\quad&\left(T_{g+2}^{H}-\frac{1}{m-1}T_{g+1}^{H}\right)-\left(T_{g+1}^{H}-\frac{1}{m-1}T_{g}^{H}\right)\nonumber \\
&=&\frac{1}{m-1}\left(T_{g+1}^{P}-T_{g}^{P}\right)=\frac{1}{m-1}\left(1+T_{g}^{H}\right),
\end{eqnarray}
where the relation provided in Eq.~(\ref{MTTa05}) was used. Applying the initial condition $T_{2}^{H}=\frac{2m-1}{m-1}$, we solve Eq.~(\ref{MTTa07}) to obtain
\begin{equation}\label{Appa08}
T_{g}^H=\left(3-\frac{5m-2}{m^2}\right)\left(\frac{m}{m-1}\right)^{g}-1.
\end{equation}
Inserting the result for $T_{g}^H$ into Eq.~(\ref{MTTa05}), we arrive the exact formula for $T_{g}^P$ given by
\begin{equation}\label{Appa09}
T_{g}^{P}=\left(3m-8+\frac{7m-2}{m^{2}}\right)\left(\frac{m}{m-1}\right)^{g}-2m+3.
\end{equation}

\section{Derivation of $R_{g}(n)$ and $H_{g}(n)$ \label{App02}}

For the two quantities $R_{g}(n)$ and $H_{g}(n)$, the following relations hold:
\begin{eqnarray}\label{MTTc03}
R_g(n)&=&\frac{1} { K_{g-n}^{R}}\big \{(m-2)[1+R_g(n)]+ \nonumber \\
&\quad& [1+H_g(n-1)+R_g(n)]+1+[2+R_g(n)]
\nonumber \\ &\quad& +\sum_{i=2}^{g-(n+2)}[1+T_{i}^{H}+R_{g}(n)]\big \}\,
\end{eqnarray}
and
\begin{eqnarray}\label{MTTc04}
H_{g}(n)&=&\frac{1}{K_{g-n}^{H}}\big \{ (m-1)^{g-n}[1+R_g(n-1)+H_g(n)]+ \nonumber \\ &\quad& (m-1)^{g-(n+1)}+(m-1)[m+H_g(n)] \nonumber \\ &\quad& +\sum_{i=2}^{g-(n+2)}(m-1)^{i}[1+T_{i}^{P}+H_g(n)]\big \} \,.
\end{eqnarray}

Equation~(\ref{MTTc03}) can be elaborated as follows. Originating from a node in $R_{g-n}$, the particle can jump to one of the $m-2$ neighboring nodes belonging to $R_{g-n}$, from which it continues to jump $R_g(n)$ steps to first visit a target; this is accounted for by the first term on the rhs. Alternatively, the walker can go to a local hub belonging to $H_{g-(n-1)}$, then takes time $H_g(n-1)$ to reach a neighbor in $R_{g-n}$, and proceeds to bounce $R_g(n)$ steps to hit a target for the first time, this process is explained by the second term. The third term describes the process by which the walker goes directly to a target node. The fourth term represents the process that the walker first jumps to a neighbor belonging to $H_{g-(g-1)}$, makes a move returning to a node in $R_{g-n}$, and then walks continuously in time $R_g(n)$ to arrive at a destination node. Finally, the last sum term explains the fact that the particle goes to a local hub in $\mathbb{H}_{g-i}$ ($2 \leq i \leq g-(n+2)$) from which it takes an average time $T_{i}^{H}$ to return to one of its neighbors in $R_{g-n}$, and then moves on average $R_{g-n}$ steps to get to a target. Analogously, we can explain Eq.~(\ref{MTTc04}).

After some algebra, Eqs.~(\ref{MTTc03}) and~(\ref{MTTc04}) can be simplified to
\begin{equation}\label{MTTc05}
R_{g}(n)=m-2+g-n+H_{g}(n-1)+1+\sum_{i=2}^{g-(n+2)}T_{i}^{H} \,,
\end{equation}
and
\begin{eqnarray}\label{MTTc06}
H_{g}(n)&=&(m-1)^{n+1-g}\sum_{i=1}^{g-n}(m-1)^{i}+(m-1)R_{g}(n-1)\nonumber \\ &\quad&
+(m-1)^{n+3-g}+\nonumber \\ &\quad&(m-1)^{n+1-g}\sum_{i=2}^{g-(n+2)}[(m-1)^{i}T_{i}^{P}]  \,.
\end{eqnarray}
Inserting Eq.~(\ref{MTTc06}) into Eq.~(\ref{MTTc05}) and utilizing the initial condition $R_g(0)=m-1+g+T_{g}^{H}+\sum_{i=2}^{g-2}T_{i}^{H}=\frac{(m-1)(3m-2)(m^2-m+1)}{m}\left(\frac{m}{m-1}\right)^{g} - 2m + 3$, Eq.~(\ref{MTTc06}) is solved to get
\begin{eqnarray}\label{Appb07}
R_{g}(n)&=&(m-1)^{n/2+1} \left[ \frac{3m-2}{ m } \left(\frac{m}{m-1}\right)^g - 2 \right]\nonumber \\  &\quad& - \frac{3m - 2}{m-1}  \left( \frac{m}{m-1} \right)^{g-n-3} + 1  \,.
\end{eqnarray}
Substituting this expression for $R_{g}(n)$ into Eq.~(\ref{MTTc06}) and solving Eq.~(\ref{MTTc06}), we obtain
\begin{eqnarray}\label{Appb08}
H_{g}(n)&=&(m-1)^{(n + 3)/2 } \left[  \frac{3m - 2 }{m}\left(\frac{m}{m-1}\right)^{g} - 2 \right] \nonumber \\ &\quad& -(3 m-2) \left(\frac{m}{m-1}\right)^{g-n-3} + 2m-3   \,.
\end{eqnarray}


\begin{references}

\bibitem{AlBa02}
R. Albert and A.-L. Barab\'asi, Rev. Mod. Phys. {\bf 74}, 47 (2002).

\bibitem{DoMe02}
S. N. Dorogovtsev and J. F. F. Mendes, Adv. Phys. {\bf 51}, 1079
(2002).

\bibitem{Ne03}
M. E. J. Newman, SIAM Rev. {\bf 45}, 167 (2003).

\bibitem{BoLaMoChHw06}
S. Boccaletti, V. Latora, Y. Moreno, M. Chavez and D.-U. Hwanga,
Phys. Rep. {\bf 424}, 175 (2006).

\bibitem{BaAl99} A.-L. Barab\'asi and R. Albert,
       Science {\bf 286}, 509 (1999).

\bibitem{GiNe02}
M. Girvan and M. E. J. Newman, Proc. Natl. Acad. Sci. U.S.A. {\bf
99}, 7821 (2002).

\bibitem{RaSoMoOlBa02}
E. Ravasz, A. L. Somera, D. A. Mongru. Z. N. Oltvai, and A.-L.
Barab\'asi, Science {\bf 297}, 1551 (2002).

\bibitem{RaBa03}
E. Ravasz and A.-L. Barab\'asi, Phys. Rev. E {\bf 67}, 026112
(2003).

\bibitem{ChLu02}
 F. Chung and L. Lu, Proc. Natl. Acad. Sci. U.S.A. {\bf 99}, 15879 (2002).

\bibitem{CoHa03}
R. Cohen and S. Havlin, Phys. Rev. Lett. {\bf 90}, 058701 (2003).

\bibitem{DaDuDiAr05}
L. Danon, J. Duch, A. Diaz-Guilera, and A. Arenas, J. Stat. Mech.:
Theory Exp. (2005) P09008.

\bibitem{PaDeFaVi05}
G. Palla, I. Der\'enyi, I. Farkas, and T. Vicsek, Nature (London)
{\bf 435}, 814 (2005).

\bibitem{Ne06}
M. E. J. Newman, Proc. Natl. Acad. Sci. U.S.A. {\bf 103}, 8577
(2006).

\bibitem{Fo10}
S. Fortunato, Phys. Rep. {\bf 486}, 75 (2010).

\bibitem{ZhZhZhGu11}
W. H. Zhan, Z. Z. Zhang, J. H. Guan, and S. G. Zhou, Phys.
Rev. E {\bf 83}, 066120 (2011).

\bibitem{Pe01}
C. E. Perkins, \emph{Ad Hoc Networking}, 1st ed. (Addison-Wesley,
Reading, MA, 2001).

\bibitem{WuGaHa04}
A. Y. Wu, M. Garland, and J. Han,
in \emph{KDD '04: Proceedings of the Tenth ACM SIGKDD International
Conference on Knowledge Discovery and Data Mining} (ACM, New
York, 2004), pp. 719-724.

\bibitem{DoGoMe08}
S. N. Dorogovtsev, A. V. Goltsev and J. F. F. Mendes,
       Rev. Mod. Phys. {\bf 80}, 1275 (2008).

\bibitem{HaBe87}
S. Havlin and D. ben-Avraham, Adv. Phys. {\bf 36}, 695 (1987).

\bibitem{MeKl00}
R. Metzler and J. Klafter, Phys. Rep. {\bf 339}, 1 (2000).

\bibitem{MeKl04}
R. Metzler and J. Klafter, J. Phys. A {\bf 37}, R161 (2004).

\bibitem{BuCa05}
R Burioni and D Cassi, J. Phys. A {\bf 38}, R45 (2005).

\bibitem{SoMaBl97}
I. M. Sokolov, J. Mai, and A. Blumen, Phys. Rev. Lett. {\bf 79}, 857
(1997).

\bibitem{PaAm04}
S. A. Pandit and R. E. Amritkar, Phys. Rev. E {\bf 63}, 041104
(2001).

\bibitem{NoRi04}
J. D. Noh and H. Rieger, Phys. Rev. Lett. {\bf 92}, 118701 (2004).

\bibitem{BeCoMoSuVo05}
O. B\'enichou, M. Coppey, M. Moreau, P.-H. Suet, and R. Voituriez,
Phys. Rev. Lett. {\bf 94}, 198101 (2005).

\bibitem{SoRebe05}
V. Sood, S. Redner, and D. ben-Avraham, J. Phys. A {\bf 38}, 109
(2005).

\bibitem{Bobe05}
E. M. Bollt and D. ben-Avraham, New J. Phys. {\bf 7}, 26 (2005).

\bibitem{GaSoHaMa07}
L. K. Gallos, C. Song, S. Havlin, and H. A. Makse, Proc. Natl. Acad.
Sci. U.S.A. {\bf 104}, 7746 (2007).

\bibitem{BaBeWi08}
P. Barthelemy, J. Bertolotti, and D. S. Wiersma, Nature (London)
{\bf 453}, 495 (2008).

\bibitem{BaCaPa08}
A. Baronchelli, M. Catanzaro, and R. Pastor-Satorras, Phys. Rev. E
{\bf 78}, 011114 (2008).

\bibitem{KiCaHaAr08}
A. Kittas, S. Carmi, S. Havlin, and P. Argyrakis, EPL {\bf 84},
40008 (2008).

\bibitem{ZhZhZhYiGu09}
Z. Z. Zhang, Y. C. Zhang, S. G. Zhou, M. Yin, and J. H. Guan, J.
Math. Phys. {\bf 50}, 033514 (2009).

\bibitem{HaRo09}
C. P. Haynes and A. P. Roberts, Phys. Rev. Lett. {\bf 103}, 020601
(2009).

\bibitem{ZhAlHoZhCh11}
Z. Z. Zhang, J. L. T. Alafate, B. Y. Hou, H. J. Zhang, and G. R. Chen, Eur. Phys.
J. B {\bf 84}, 691 (2011).

\bibitem{We94}
G. H. Weiss, \emph{Aspects and Applications of the Random Walk}
(North Holland, Amsterdam, 1994).

\bibitem{Hu95}
B. H. Hughes, \emph{Random Walks and Random Environments} (Clarendon
Press, Oxford, 1995), Vols. 1 and 2.

\bibitem{JaBl01}
F. Jasch and A. Blumen, Phys. Rev. E {\bf 63}, 041108 (2001).

\bibitem{Sh05}
M. F. Shlesinger, Nature (London) {\bf 443}, 281 (2006).

\bibitem{WhGo99}
J. Whitmarsh and J. A. Govindjee, in \emph{Concepts in Photobiology:
Photosynthesis and Photomorphogenesis}, edited by G. Singhal, G.
Renger, S. Sopory, K.-D. Irrgang, and J. A. Govindjee (Narosa, New
Delhi, 1999), pp. 11-51.

\bibitem{FoPiReSa07}
F. Fouss, A. Pirotte, J. M. Renders, and M. Saerens, IEEE Trans.
Knowl. Data Eng. {\bf 19}, 355 (2007).

\bibitem{Mo69}
E. W. Montroll, J. Math. Phys. {\bf 10}, 753 (1969).

\bibitem{KaBa02PRE}
J. J. Kozak and V. Balakrishnan, Phys. Rev. E {\bf 65}, 021105
(2002).

\bibitem{KaBa02IJBC}
J. J. Kozak and V. Balakrishnan, Int. J. Bifurcation Chaos Appl.
Sci. Eng. {\bf 12}, 2379 (2002).


\bibitem{ZhWuZhZhGuWa10}
Z. Z. Zhang,  B. Wu, H. J. Zhang, S. G. Zhou, J. H. Guan, and Z. G.
Wang, Phys. Rev. E {\bf 81}, 031118 (2010).

\bibitem{Ag08}
E. Agliari, Phys. Rev. E {\bf 77}, 011128 (2008).

\bibitem{ZhLiZhWuGu09}
Z. Z. Zhang, Y. Lin, S. G. Zhou, B. Wu, and J. H. Guan, New J. Phys.
{\bf 11}, 103043 (2009).

\bibitem{CaAb08}
A. Garcia Cant\'u and E. Abad, Phys. Rev. E {\bf 77}, 031121 (2008).

\bibitem{ZhQiZhXiGu09}
Z. Z. Zhang, Y. Qi, S. G. Zhou, W. L. Xie, and J. H. Guan, Phys.
Rev. E {\bf 79}, 021127 (2009).

\bibitem{ZhZhXiChLiGu09}
Z. Z. Zhang,  S. G. Zhou, W. L. Xie, L. C. Chen, Y. Lin,  and J. H.
Guan, Phys. Rev. E {\bf 79}, 061113 (2009).

\bibitem{ZhGuXiQiZh09}
Z. Z. Zhang, J. H. Guan, W. L. Xie, Y. Qi, and S. G. Zhou, EPL, {\bf
86}, 10006 (2009).

\bibitem{AgBu09}
E. Agliari and R. Burioni, Phys. Rev. E {\bf 80}, 031125 (2009).

\bibitem{TeBeVo09}
V. Tejedor, O. B\'enichou, and R. Voituriez, Phys. Rev. E {\bf 80},
065104(R) (2009).

\bibitem{ZhXiZhGaGu09}
Z. Z. Zhang, W. L. Xie, S. G. Zhou, S. Y. Gao, and J. H. Guan, EPL
{\bf 88}, 10001 (2009).

\bibitem{ZhXiZhLiGu09}
Z. Z. Zhang, W. L. Xie, S. G. Zhou, M. Li, and J. H. Guan, Phys.
Rev. E {\bf 80}, 061111 (2009).


\bibitem{CoBeTeVoKl07}
S. Condamin, O. B\'enichou, V. Tejedor, R. Voituriez, and J.
Klafter, Nature (London) {\bf 450}, 77 (2007).

\bibitem{BeMeTeVo08}
O. B\'enichou, B. Meyer, V. Tejedor, and R. Voituriez, Phys. Rev.
Lett. {\bf 101}, 130601 (2008).

\bibitem{CoTeVoBeKl08}
S. Condamin, V. Tejedor, R. Voituriez, O. B\'enichou and J. Klafter,
Proc. Natl. Acad. Sci. U.S.A. {\bf 105}, 5675 (2008).

\bibitem{YuAc03}
S. B. Yuste and L. Acedo, Phys. Rev. E {\bf 68}, 036134 (2003).

\bibitem{WiSa82}
T. A. Witten and  L. M. Sander, Phys. Rev. Lett. {\bf 47}, 1400
(1981).

\bibitem{WiSa83}
T. A. Witten and  L. M. Sander, Phys. Rev. B {\bf 27}, 5686 (1983).

\bibitem{Ri85}
S. A. Rice, \emph{Diffusion-Limited Reactions}, edited by C. H. Bamford, C. F. H. Tipper, and R. G. Compton, Comprehensive
Chemical Kinetics Vol. 25 (Elsevier, Amsterdam, 1985).

\bibitem{Hi07}
M. Hinczewski, Phys. Rev. E {\bf 75}, 061104 (2007).

\bibitem{Ma07}
S. Maslov, Nature Phys. {\bf 3}, 18 (2007).  

\bibitem{Di08}
A. Diaz-Guilera, J. Phys. A {\bf 41}, 224007 (2008).

\bibitem{ZhLiGoZhGuLi09}
Z. Z. Zhang, Y. Lin, S. Y. Gao, S. G. Zhou, J. H. Guan, and M. Li,
Phys. Rev. E {\bf 80}, 051120 (2009).

\bibitem{BuSp09}
E. Bullmore and O. Sporns, Nature Rev. Neurosci. {\bf 10}, 186 (2009).

\bibitem{Re01}
S. Redner, \emph{A Guide to First-Passage Processes} (Cambridge
University Press, Cambridge, 2001).

\bibitem{DoGoMe02}
S. N. Dorogovtsev, A. V. Goltsev, and J. F. F. Mendes,
          Phys. Rev. E {\bf 65}, 066122 (2002).

\bibitem{JuKiKa02}
S. Jung, S. Kim, and B. Kahng,
        Phys. Rev. E {\bf 65}, 056101 (2002).

\bibitem{CoOzPe00}
F. Comellas, J. Oz\'on, and J.G. Peters, Inf. Process. Lett. {\bf
76}, 83 (2000)

\bibitem{ZhRoGo06}
Z. Z. Zhang, L. L Rong, and C. H. Guo, Physica A {\bf 363}, 567
(2006).

\bibitem{HiBe06}
M. Hinczewski and A. N. Berker, Phys. Rev. E {\bf 73}, 066126
(2006).

\bibitem{RoHaAv07}
H. D. Rozenfeld, S. Havlin, and D. ben-Avraham, New J. Phys. {\bf
9}, 175 (2007).


\bibitem{ZhZhChGu07}
Z. Z. Zhang, S. G. Zhou, T. Zou, L. C. Chen, and J. H. Guan, Eur.
Phys. J. B {\bf 60}, 259 (2007).

\bibitem{ZhZhFaGuZh07}
Z. Z. Zhang, S. G. Zhou, L. J. Fang, J. H. Guan, and Y. C. Zhang,
EPL {\bf 79}, 38007 (2007).

\bibitem{BoGoGu08}
S. Boettcher, B. Gon\c calves, and H. Guclu, J. Phys. A {\bf 41},
252001 (2008).

\bibitem{CoMi09}
F. Comellas and A. Miralles, Physica A. {\bf 388}, 2227 (2009).

\bibitem{MoRoDa09}
K. L. Morrow, T. Rowland, and C. M. Danforth, Phys. Rev. E {\bf 80},
016103 (2009).

\bibitem{ZhGuDiChZh09}
Z. Z. Zhang, J. H. Guan, B. L. Ding, L. C. Chen, and S. G. Zhou, New
J. Phys. {\bf 11}, 083007 (2009).

\bibitem{MiCoChZh10}
A. Miralles, F. Comellas, L. C. Chen, and Z. Z. Zhang, Physica A.
{\bf 389}, 1955 (2010).

\bibitem{No03}
J. D. Noh, Phys. Rev. E {\bf 67}, 045103(R) (2003).

\bibitem{NoRi04a}
J. D. Noh and H. Rieger, Phys. Rev. E {\bf 69}, 036111 (2004).


\bibitem{BaRaVi01}
A.-L. Barab\'asi, E. Ravasz, and T. Vicsek,
          Physica A  {\bf 299}, 559 (2001).

\bibitem{IgYa05}
K. Iguchi and H. Yamada, Phys. Rev. E {\bf 71}, 036144 (2005).

\bibitem{ZhLiGaZhGu09}
Z. Z. Zhang, Y. Lin, S. Y. Gao, S. G. Zhou,  and J. H. Guan, J.
Stat. Mech. (2009) P10022.


\bibitem{AgBuMa10}
E. Agliari, R. Burioni, and A. Manzotti, Phys. Rev. E {\bf 82},
011118 (2010).

\bibitem{ZhLi10}
Z. Z. Zhang and Y. Lin, J. Stat. Mech. (2010) P12017.


\bibitem{WaSt98} D. J. Watts and H. Strogatz,
        Nature (London) {\bf 393}, 440 (1998).

\bibitem{KeSn76}
J. G. Kemeny and J. L. Snell, \emph{Finite Markov Chains} (Springer,
New York, 1976).

\bibitem{AlFi99}
D. Aldous and J. Fill, http://www.stat.berkeley.edu/~aldous/RWG/Chap2.pdf

\bibitem{BaKl98}
A. Bar-Haim and J. Klafter, J. Chem. Phys. {\bf 109}, 5187 (1998).

\bibitem{note}
Note that the definition of MFPT $\langle T \rangle$ slightly differs from that used in some previous work where the average of $ T_i$ is taken over all nodes other than the traps. However, when the concentration of traps (the ratio of the trap number $\Gamma$ to network size $N$) is very small, both definitions lead to the same behavior of MFPT $\langle T \rangle$.


\bibitem{Wi94}
H. S. Wilf, \emph{Generatingfunctionology}, 2nd ed. (Academic,
London, 1994).


\bibitem{TeBeVo11}
V. Tejedor, O. B\'enichou, and R. Voituriez, Phys. Rev. E {\bf 83},
066102 (2011).



\end{references}
\end{document}